\documentclass[12pt]{article}

\newread\testifexists
\def\GetIfExists #1 {\immediate\openin\testifexists=#1
    \ifeof\testifexists\immediate\closein\testifexists\else
    \immediate\closein\testifexists\input #1\fi}

\usepackage{gthstyle}
\immediate\openin\testifexists=e
    \ifeof\testifexists\immediate\closein\testifexists\else
    \immediate\closein\testifexists\input e\fipsf

\def\Bbb#1{\setbox0=\hbox{$\tt #1$}  \copy0\kern-\wd0\kern .1em\copy0}

\def\bbf#1{\setbox0=\hbox{$#1$} \kern-.025em\copy0\kern-\wd0
        \kern.05em\copy0\kern-\wd0 \kern-.025em\raise.0433em\box0}

\immediate\openin\testifexists=a
    \ifeof\testifexists\immediate\closein\testifexists\else
    \immediate\closein\testifexists\input a\fimssym.def          

      \def\b{\beta}         \def\G{\Gamma}
        \def\e{\varepsilon}
               
\def\m{\mu}

 \def\ra{\rightarrow}

\def\ffract#1#2{\raise .3 em\hbox{$\scriptstyle#1$}\kern-.25em/
                \kern-.2em\lower .2 em \hbox{$\scriptstyle#2$}}

\def\part#1#2{{\partial#1\over\partial#2}}

\newcommand{\be}{\begin{eqnarray}}
\newcommand{\lee}[1]{\label{#1}\end{eqnarray}}
\newcommand{\ee}{\end{eqnarray}}
\newcommand{\eqn}[1]{(\ref{#1})}
\newcommand{\nnn}{\nonumber\\}

\newcommand{\rf}[1]{\cite{#1}}

\newcommand{\fn}{\footnote}

\newcommand{\newsec}[1]{\section{#1}\setcounter{equation}{0}}

\begin{document}

\begin{titlepage}

\title{\normalsize \hfill ITP-UU-04/13  \\ \hfill SPIN-04/07
\\ \hfill {\tt hep-th/0405032}\\ \vskip 20mm \Large\bf
RENORMALIZATION WITHOUT INFINITIES
\thanks{Presented at the \emph{Coral Gables Conference,
Fort Lauderdale, Fa, Dec. 16-21, 2003}}
\author{Gerard 't~Hooft}
\date{\normalsize Institute for Theoretical Physics \\
Utrecht University, Leuvenlaan 4\\ 3584 CC Utrecht, the
Netherlands\medskip \\ and
\medskip \\ Spinoza Institute \\ Postbox 80.195 \\ 3508 TD
Utrecht, the Netherlands \smallskip \\ e-mail: \tt
g.thooft@phys.uu.nl \\ internet: \tt
http://www.phys.uu.nl/\~{}thooft/}}  \maketitle

\begin{quotation} \noindent {\large\bf Astract } \medskip \\
Most renormalizable quantum field theories can be rephrased in
terms of Feynman diagrams that only contain dressed irreducible
2-, 3-, and 4-point vertices. These irreducible vertices in turn
can be solved from equations that also only contain dressed
irreducible vertices. The diagrams and equations that one ends up
with do not contain any ultraviolet divergences. The original bare
Lagrangian of the theory only enters in terms of freely adjustable
integration constants. It is explained how the procedure proposed
here is related to the renormalization group equations. The
procedure requires the identification of unambiguous ``paths" in
a Feynman diagrams, and it is shown how to define such paths in
most of the quantum field theories that are in use today. We do
not claim to have a more convenient calculational scheme here,
but rather a scheme that allows for a better conceptual
understanding of ultraviolet infinities.
\end{quotation}

\vfill \flushleft{\today}

\end{titlepage}


{\large\textbf{Congratulations}}\bigskip

\noindent This contribution is written at the occasion of Paul
Frampton's \(60^{\underline{\mathrm{th}}}\) birthday.
\newsec{Introduction. Rearranging Feynman diagrams}

Usually, a quantized field theory is defined through its bare
Lagrangian. From this Lagrangian, one derives Feynman diagrams to
represent contributions to the amplitudes that one wishes to
compute. Many of the resulting expressions are found to contain
ultraviolet divergences, which are subsequently neutralized by
adding new counter terms to the original Lagrangian\rf{IZ}. In
practice, this works so well that refinements and causeats are not
thought worth-while for consideration, and indeed one can
formulate precise justifications of this procedure. If the total
number of physically distinguishable freely adjustable parameters
stays finite and fixed in the course of the perturbative
expansion, the theory is called `renormalizable'.

In many cases, however, there exists an alternative way to
address renormalizable theories, such that never in the
intermediate results UV divergent expressions enter. This short
communication aims at explaining it.

\begin{figure}[ht]\setcounter{figure}{0}
\begin{center} \epsfxsize=85 mm\epsfbox{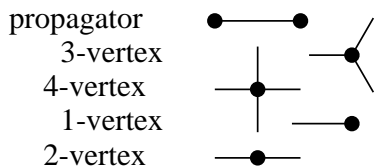}
  \caption{\small{Bare propagators and vertices\label{vertices.fig}
}}
\end{center}
\end{figure}
We start with the original formulation of the Feynman rules. In
general, there are three-point vertices and four-point vertices.
Higher vertices will rend the theory unrenormalizable, with the
exception of some theories in lower space-time dimensions which
we shall not consider.

Vertices with one or two external lines may occur, but they can
quickly be eliminated by shifting and renormalizing the field
definitions, and hence we will ignore them, although with a
little more effort one can accommodate for them in the
formulation below (see Fig.~\ref{vertices.fig}).

\begin{figure}[ht]
\begin{center} \epsfxsize=160 mm\epsfbox{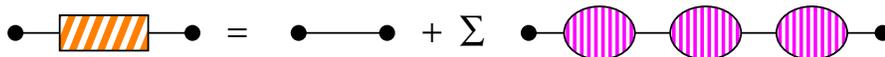}
  \caption{\small{The dressed propagator.}\label{dressedprop.fig}}
\end{center}
\end{figure}
\begin{figure}[ht]
\begin{center} \epsfxsize=140 mm\epsfbox{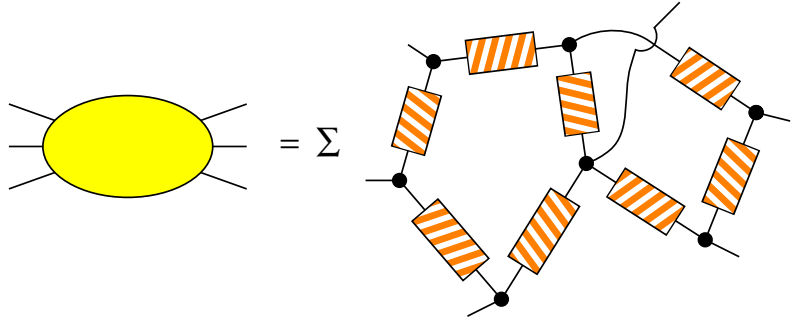}
  \caption{\small{Example of a diagram with dressed propagator
  insertions}
  \label{drpropindiagr.fig}}
\end{center}
\end{figure}

Our procedure for rearranging Feynman diagrams begins with
dressing all propagators. We use the notation illustrated in
Fig.~\ref{dressedprop.fig}. Thus, from now on, all propagators in
a diagram are assumed to include the one-particle irreducible
2-point diagrams, which form a geometric series. A generic
diagram will look as illustrated in Fig.~\ref{drpropindiagr.fig}.

\begin{figure}[ht]
\begin{center} \epsfxsize=120 mm\epsfbox{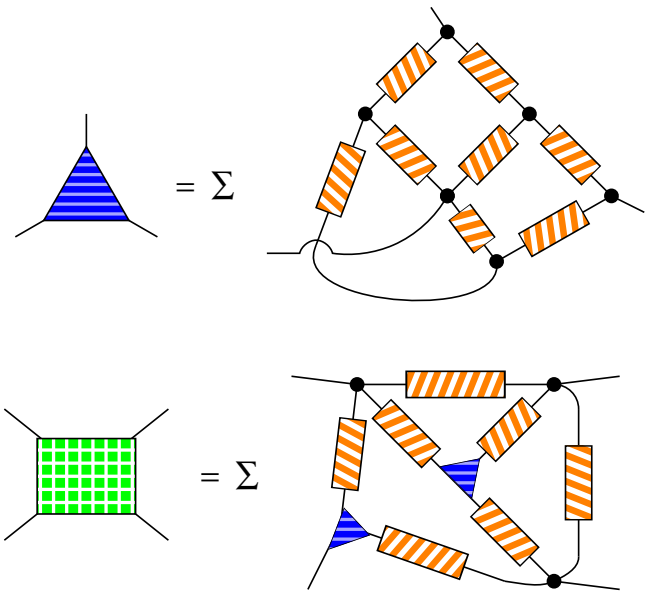}
  \caption{\small{The dressed 3-point vertices formed from irrdeucible
  subdiagrams, and the dressed 4-point vertices from
  irrdeucible 4-point diagrams}\label{34irreducible.fig}}
\end{center}
\end{figure}

\begin{figure}[ht]
\begin{center} \epsfxsize=130 mm\epsfbox{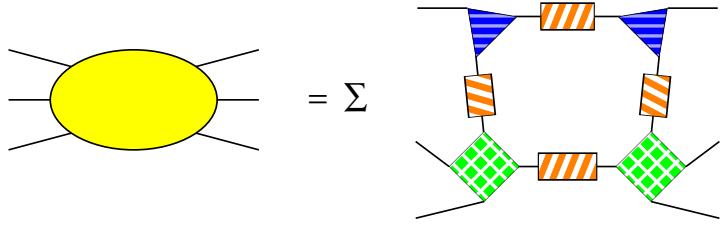}
  \caption{\small{Diagrams with more than 4 external lines are built
  exclusively of dressed propagators, dressed 3-point vertices
  and dressed 4-point vertices}\label{dr34indiagr.fig}}
\end{center}
\end{figure}

Next, we consider all one-particle irreducuble 3-point diagrams.
They can also be added, once and for all, to all bare 3-point
vertices, to for m the so-called dressed 3-point vertices.
Similarly, we can collect all subdiagrams needed to turn all
4-vertices into dressed 4-vertices, see
Fig.~\ref{34irreducible.fig}. It is important that diagrams, where
the propagators and vertices are replaced by dressed ones,
themselves should not contain any other subgraphs with three or
four external lines.
 A generic diagram then looks as in
Fig.~\ref{dr34indiagr.fig}.

\begin{figure}[ht]
\begin{center} \epsfxsize=100 mm\epsfbox{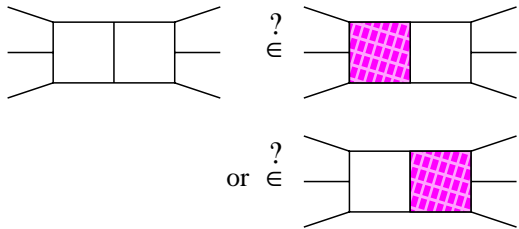}
  \caption{\small{Ambiguity for irreducible 5-point vertices: the diagram
  indicated has six external lines and can be written in two ways
  using an irreducible 5-vertex. It would be counted twice, which is
  incorrect unless further correction procedures are introduced}
  \label{overcount.fig}}
\end{center}
\end{figure}

\begin{figure}[ht]
\begin{center} \epsfxsize=120 mm\epsfbox{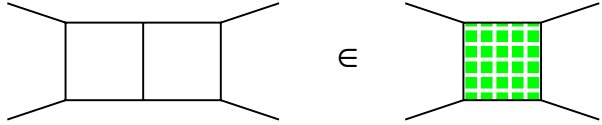}
  \caption{\small{Subgraphs with 4 external lines do not lead to counting errors.
  The diagram depicted here contributes just once to the irreducble
  dressed 4-vertex}\label{fourcount.fig}}
\end{center}
\end{figure}

It is important to check here that rearranging diagrams using
this prescription does not lead to omissions of any diagrams or
to overcounting of diagrams. Indeed, if we were to continue the
procedure towards irreducible diagrams with \emph{five} external
lines, overcounting would occur. This is illustrated in
Fig.~\ref{overcount.fig}. Such an ambiguity cannot occur in the
case of 4-vertices; cf. Fig.~\ref{fourcount.fig}. The diagram of
this Figure is counted correctly as a single contribution to the
dressed 4-vertices.

We conclude from this section that all diagrams with five or more
external lines can be seen to be built up in an unambiguous way
from irreducible dressed propagators, 3-point functions and
4=point functions. These dressed diagrams themselves should not
contain any irreducible subgraph with less than five external
lines. Consequently, the integrations over any of the momenta in
these dressed diagrams do not lead to any ultraviolet divergence.
In particular, there are no overlapping ultraviolet divergences.

However, the dressed 2-, 3- and 4-point functions themselves
cannot be reduced to convergent integrals along such lines; they
themselves still seem to be built out of \emph{bare} propagators
and vertices. They will be considered in the next sections.

\newsec{The Ariadne Procedure}

The dressed 2-, 3- and 4-point vertices may be
divergent\fn{Diagrams with \emph{fermions} are less divergent;
one may decide to count external fermions with weight $3/2$ in
the procedure that follows.}, but if we introduce
\emph{subtractions}, more convergent expressions may arise. We
claim that, if a divergent, irreducible diagram with $n<5$
external lines is considered, then we can take the
\emph{difference} between that diagram and the same diagram at
some different values of its external momenta, and rewrite that
as a new irreducible Feynman diagram with $n+1$ external lines,
whose degree of convergence is improved by at least one power of
$k$.

\begin{figure}[ht]
\begin{center} \epsfxsize=70 mm\epsfbox{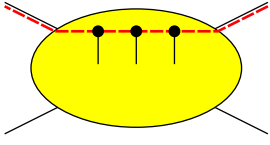}
  \caption{\small{The introduction of a guiding path (dotted line)
  }\label{path.fig}}
\end{center}
\end{figure}

In order to introduce unambiguous rules for these \emph{difference
diagrams}, we need the notion of a guiding path inside a
diagram.\footnote{In planar diagram theories, the guiding path is
simply the edge of a diagram. In fact, this was used in
Ref\rf{GtHPlanar}.} A guiding path is a sequence of propagators
inside a diagram that form a single uninterrupted line from one
external line to another, see Fig.~\ref{path.fig}. If an external
line is a fermion, such as in QED, we can use this fermion as a
guiding path. In $U(N)$ (gauge) theories, we can often use index
lines as guiding paths, provided that not all index lines lead
from one external line back to the same one; since such diagrams
do not contribute in $SU(N)$ theories (fields in the adjoint
representation are traceless), index lines are assured to be
useful in these theories. However, also in $SO(N)$ theories, index
lines cannot run from one external line back to the same line,
since here also the adjoint representation is traceless. This
means that also $U(1)$ theories can sometimes be handled. If,
however, an external line has two (or more) units of $U(1)$
charge, it means that it has two index lines in terms of which the
representation is symmetric, hence not traceless. In that case, we
have to use some other guiding line. In the electro-weak case,
this appears to be possible: all our bosonic fields have several
quantum numbers, so that we do not have to resort on the
unreliable $U(1)$ indices.

Thus, it seems that in most cases of interest one can find a
guiding path. This will be referred to as the Ariadne principle.
It is the one restriction that we will assume, besides the more
familiar restriction that our theory should not contain any chiral
anomalies\rf{ABJ} (more about the anomalies later).

\begin{figure}[ht]
\begin{center} \epsfxsize=140 mm\epsfbox{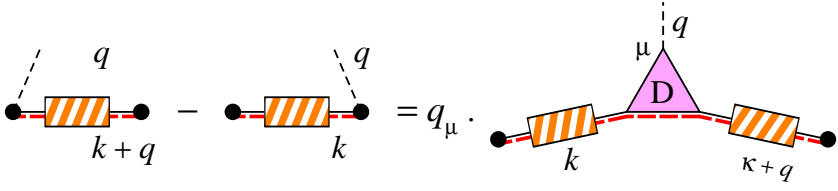}
  \caption{\small{Difference between two dressed propagators
  in diagrammatic notation}\label{prpdiff.fig}}
\end{center}
\end{figure}

Consider the sequence of (bare) propagators $P(k_i)$ and (bare)
vertices $V(k_i)$ along a guiding path that leads from an external
line with momentum $k_0$ to another line with momentum $k_N$. \be
F(k_0,\,k_N)=V_1(k_1)P_1(k_1)\,\cdots\,
P(k_{N-1})V_N(k_N)\,;\lee{pathf} Here, and $k_i$ is the momentum
in the $i^\mathrm{th}$ propagator or vertex, $i$ running from 1 to
$N$. Now substitute all these momenta by the same values plus an
additional, fixed, momentum $q$, and compute the difference
between these two amplitudes: \be &&F(k_0+q,\,k_N+q)-F(k_0,\,k_N)\
= \sum_i\Bigg\{V_1(k_1)\cdots P_{i-1}(k_{i-1})\cr &&
\bigg(\Big(V_i(k_i+q)-V_i(k_i)\Big)P_i(k_i+q)+V_i(k_i)\Big(P_i(k_i+q)-P_i(k_i)\Big)\bigg)\cr
&&V_{i+1}(k_{i+1}+q)\,P_{i+1}(k_{i+1}+q)\,\cdots\,
V_N(k_N+q)\Bigg\}\,.\lee{pathdiff} We see that this expression
contains bare propagators and vertices that again form dressed
propagators and vertices when summed. In particular, parts of this
expression refer to the difference between two dressed
propagators, which obey \def\dr{\mathrm{\,dr}} \be
P^\dr(k)&=&\Big(k^2+m^2-\G(k)-i\e\Big)^{-1}\quad; \nnn P^\dr(k+q)
- P^\dr(k)&=&
P^\dr(k)\,\Big[\G(k+q)-\G(k)-(k+q)^2+k^2\Big]\,P^\dr(k+q)\,.\
{}\lee{propdiff} Writing \be k^2-(k+q)^2=-q_\m(2k^\m+q^\m)\
;\qquad\G(k+q)-\G(k)=q_\m\G_1^\m(k,q)\ , \lee{factq} we see that
the expression between square brackets here can be regarded as an
effective three-point diagram \(\G_1^\m\), multiplied with a
factor \(q_\m\). Eq.~\eqn{propdiff} is depicted diagrammatically
in Fig.~\ref{prpdiff.fig}.

\newsec{Integrating the equations. Conclusion}

\begin{figure}[ht]
\begin{center} \epsfxsize=130 mm\epsfbox{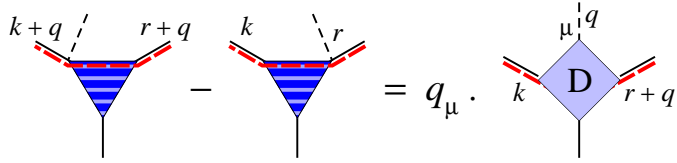}
  \caption{\small{The difference between the 3-point
  functions at different values of the external momenta is a
  four-point diagram}\label{threediff.fig}}
\end{center}
\end{figure}

\begin{figure}[ht]
\begin{center} \epsfxsize=130 mm\epsfbox{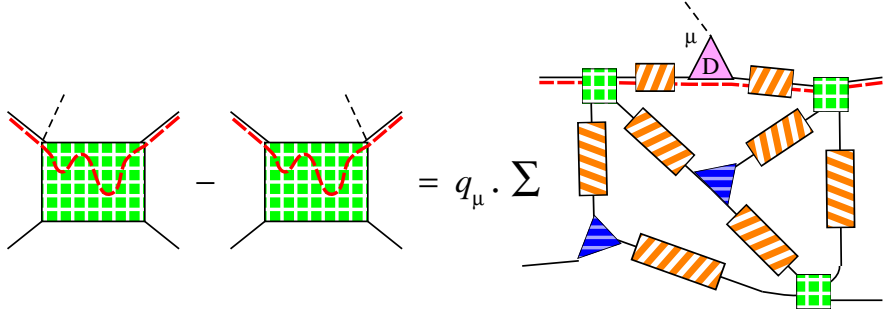}
  \caption{\small{The difference between the 4-point
  functions at different values of the external momenta is a
  sum of convergent diagrams}\label{fourdiff.fig}}
\end{center}
\end{figure}

Similarly, we can handle the difference between three-point
diagrams, $V^{[3]}(k+q)-V^{[3]}(k)$, at different external
momenta as four-point diagrams, see Fig.~\ref{threediff.fig}. The
difference between two four-point diagrams is a sum of convergent
diagrams, see Fig.~\ref{fourdiff.fig}. The difference equations
used here can be used either for the original irreducible diagrams
for the theory or for the diagrams obtained after a previous
differenciation. In all cases, the irreducible diagrams of five
or more external lines only contain convergent expressions. As
far as the ultra-violet divergences is concerned, the situation
is the same as if we had differentiated with respect to the
momenta rather than taking finite differences (\emph{i.e.}, if
$q$ had been taken infinitesimal. A disadvantage of infinitesimal
$q$, however, is the emergence of higher order poles and the
associated infra-red divergences in the propagators. Our
difference procedure avoids infra-red divergences.

Fig.~\ref{fourdiff.fig}, whose equation reads as: \be
V^{[4]}(k+q)-V^{[4]}(k)=q_\m \sum_i V^{\m\,[5]}_i (k,q,\cdots)\ ,
\lee{fourdif} where $V_i^{\m[5]}$ do not contain any divergences,
in many respects is to be regarded as a renormalization group
equation. Since $V_i^{\m[5]}$ are all linearly (or better)
convergent, the equation can be symbolized as \be {d\over d
k}\G^{[4]}(k)={1\over k}\b(\G^{[4]})(k)\,,\lee{beta} the r.h.s.
being essentially a beta function. The difference equations for
the 2- and 3-point functions, in short-hand, are\be
\G^{[2]}(k+q)-\G^{[2]}(k)&=&q^\m \,k\,(D\G)^{[3]}_\m(k,q)\ ,\label{dif2} \\
V^{[3]}(k+q)-V^{[3]}(k)&=&q^\m \,(DV)^{[4]}_\m (k,q)\ ,\lee{dif3}
where we explicitly indicated the $k$-dependence apart from
logarithms. Thus, we see that Eqs.~\eqn{dif2} and \eqn{dif3}
converge in the infrared when integrated, whereas \eqn{beta} has
the infra-red structure of the renormalization group.\rf{CS}

Our set of equations appears to be particularly elegant because
no direct reference is made to the \emph{bare} Lagrangian of the
theory! All bare coupling parameters are generated by the
integration constants when integrating these difference equations.

Thus, Quantum Field Theory has been recast into a self-consistent
set of equations, which can be integrated to obtain the desired
amplitudes.  The four-point amplitudes \linebreak[2]--~more
precisely, the canonically dimensionless irreducible $n$-point
functions~-- follow from solving the renormalization group
equation Eq.~\eqn{fourdif}. The required integration constant(s)
replace the original free parameters of the theory.

A difficulty may arise from the fact that the `guiding lines' may
be chosen in many alternative ways. Indeed, it is in integrating
the equations that one might encounter anomalies\rf{ABJ}: the
integration constants cannot be reconciled with all symmetries of
the theory.

Also, the lower irreducible Green functions may generate
integration constants, which would correspond to dimensionful
parameters of the theory. The usual questions concerning
``naturalness" are not affected by our procedure; if the
integration constants lead to small amplitudes in the far
infra-red, this may be considered as `unnatural', but there is no
objection to that from a purely mathematical point of view.

Another fundamental difficulty not addressed by our procedure is
the divergence of the perturbative expansion for the diagrams in
the r.h.s. of Eq.~\eqn{fourdif}, depicted in
Fig.~\ref{fourdiff.fig}. In general, such expansions diverge
factorially. In the planar $N\ra\infty$ limit, the number of
diagrams increases by calculable power laws, but individual
diagrams may grow factorially, so that there still is no guarantee
for a finite radius of convergence.\rf{GtHPlanar} In practice, it
seems to be not unreasonable to simply cut the series off at some
given order.


\begin{thebibliography}{99}

\bibitem{IZ} C.~Itzykson and J.-B.~Zuber, Quantum Field Theory,
McGraw-Hill 1985; ISBN 0-07-032071-3; 0-07-066353-X.
\bibitem{GtHPlanar} G,~'t Hooft, \emph{Planar diagram field
theories}, in \emph{Progrss in Gauge Field Theory, NATO Adv. Study
Inst. Series}, eds. G.~'t Hooft \emph{et al}., Plenum, 1984, 271,
reprinted in G.~'t Hooft, \emph{Under the Spell of the Gauge
principle}, World Scientific, 1994, p. 378. See also: \textit{
Nucl. Phys.}  \textbf{B 72} (1974) 461.
\bibitem{ABJ} S.L.~Adler, \textit{Phys.~Rev.}~\textbf{ 177} (1969) 2426;
J.S.~Bell and R.~Jackiw,
 \textit{Nuovo Cim.}~\textbf{60A} (1969) 47; S.L.~Adler and W.A.~Bardeen,
 \textit{Phys.~Rev.}~\textbf{182} (1969) 1517;  W.A.~Bardeen,
 \textit{Phys.~Rev.}~\textbf{ 184} (1969) 1848.
\bibitem{CS} C.G.~Callan, \textit{Phys.~Rev.} \textbf{D2} (1970)
1541; K.~Symanzik, Commun.~Math.~Phys. \textbf{16} (1970) 48;
    \textit{ibid.} \textbf{18} (1970) 227.
\end{thebibliography}
\end{document}